\documentclass[runningheads]{llncs}
\usepackage[hyphens]{url}
\usepackage[hidelinks]{hyperref}
\hypersetup{breaklinks=true}
\usepackage[T1]{fontenc}  
\usepackage{cite}  
\usepackage{graphicx}
\graphicspath{{figures/}}

\begin{document}
\title{Tenant-Aware Slice Admission Control using Neural Networks-Based Policy Agent\thanks{This work has received
funding from the H2020-MSCA-ITN-2016 SPOTLIGHT project under grant number 722788.}}
\titlerunning{Tenant-Aware Slice Admission}
%
\author{Pedro Batista\inst{1,2} \and
Shah Nawaz Khan\inst{1} \and
Peter \"Ohl\'en\inst{1} \and
Aldebaro Klautau\inst{2}}
\authorrunning{P. Batista et al.}
%
\institute{Ericsson Research, Sweden\\
\email{\{pedro.batista,shah.khan,peter.ohlen\}@ericsson.com} \and
Federal University of Pará, Brazil\\
\email{aldebaro@ufpa.br}}
\maketitle              
\begin{abstract}
5G networks will provide the platform for deploying large number of tenant-associated management, control and end-user applications having different resource requirements at the infrastructure level.
In this context, the 5G infrastructure provider must optimize the infrastructure resource utilization and increase its revenue by intelligently admitting network slices that bring the most revenue to the system.
In addition, it must ensure that resources can be scaled dynamically for the deployed slices when there is a demand for them from the deployed slices.
In this paper, we present a neural networks-driven policy agent for network slice admission that learns the characteristics of the slices deployed by the network tenants from their resource requirements profile and balances the costs and benefits of slice admission against resource management and orchestration costs.
The policy agent learns to admit the most profitable slices in the network while ensuring their resource demands can be scaled elastically.
We present the system model, the policy agent architecture and results from simulation study showing an increased revenue for infra-structure provider compared to other relevant slice admission strategies.

\end{abstract}
\section{Introduction}\label{sec:introduction}
Communication networks have been continuously evolving towards an ever-increasing complexity, both in integrating new technologies and supporting new verticals.
The former requires a cross-domain and cross-technology network deployment and optimization while the latter imposes heterogeneous requirements on the network operators and the infrastructure that must support them~\cite{gomesNextStopZerotouch2018}.
5G networks, the most recent evolution of mobile communication networks, is anticipated to be platform for not only integrating new and revolutionary technologies such as Software Defined Networking (SDN) and Network Function Virtualization (NFV) but also support the requirements of new verticals through network slicing and multi-tenancy~\cite{oliva5GTRANSFORMERSlicingOrchestrating2018}.
However, with such a wide gamut of technologies being integrated and support for different verticals developed, 5G networks have become extremely complex for management and control using the traditional network practices.
One consequence of the 5G complexity is the large number of configurable parameters that exists in the network's cloud, radio access, control and management domains.
Doing the large number of possible configurations manually is bound to trigger both suboptimal configuration setups which may lead not only to service disruption and failures but also adversely affect the revenue generation capacity of the network infrastructure.
It is well recognized in this context that to handle this complexity, network automation requiring minimal human intervention will be required~\cite{gomesNextStopZerotouch2018}.
In existing networks, automation is generally an add-on feature that is mostly driven by pre-defined set of rules in specific context of a use case such as load balancing, mobility management, interference management etc.
In 5G networks however, network automation driven by machine learning and artificial intelligence is anticipated to be a core feature that will drive most of the network management and control functions in an autonomous manner.

An important feature of 5G networks to support multi-tenancy is the network slicing concept which enables the network operator or Infrastructure Provider (InP) to facilitate different service providers in the network
by providing dedicated resources~\cite{samdanisNetwork2016}.
The service providers in turn, offer revenue for the resources allocated to their deployed services in their dedicated slice.
The network slicing concept has been considered at different scales, abstraction levels and in different network segments in the context of multi-tenant 5G networks~\cite{samdanisNetwork2016,caballeroNetwork2018,zhengStatistical2018,hanMarkov2018}.
However, regardless of how network slices are defined, they are eventually mapped onto a shared network infrastructure and must be managed by InP to optimize both resource utilization and revenue generated from the deployed slices.

In this work, we focus on the network slicing concept and present a reinforcement learning-based policy agent that aims to optimize the revenue generated from deploying different service slices in the network while ensuring that the deployed services can elastically scale their resource consumption footprint when needed.
The rest of the paper is organized as follows.
Section~\ref{sec:related} provides the related work on network slicing and platforms supporting network slice deployment in the scope of virtualized 5G networks.
Section~\ref{sec:slice-admission} presents our proposed policy agent together with the system model and close-loop management architecture.
In Section~\ref{sec:evaluation},
simulation scenarios and results are presented to substantiate the increased revenue claim when compared to other relevant slice deployment strategies.
The paper is concluded in Section~\ref{sec:conclusion} with a summary and discussion of future work in this scope.

\section{Related Work}\label{sec:related}
Virtualization is at the core of 5G network architecture where cloud platform spans across the different network segments diverging from the traditional centralized cloud architecture.
A cloud-based network infrastructure inherently supports multi-tenancy and resource sharing. In such a context, there is a need for intelligent resource management to deploy network slices on the shared infrastructure.
In this section, we describe some network slice management approaches that have been considered in the literature.

Samdanis~et~al.~\cite{samdanisNetwork2016} proposes a 5G slice network management architecture.
It is focused on having many players interacting over the network and the interfaces for communication among them.
The architecture assumes a shared radio access network, that is divided in multiple domains, each one controlled by a domain manager, which are themselves managed by a higher-level network manager called 5G network slice broker with which the tenants communicate.
To operate the network, the authors explicitly cite a set of metrics that are important for slice management.
These metrics include the amount of resources allocated to a network slice such as physical resources or data rate, timing such as starting time, duration or periodicity of a request and time window, the type of resources and Quality of Service (QoS) parameters such as radio/core bearer type, prioritization, delay jitter, loss, etc.
These metrics are important for understanding what the general service requirements at a high level are, such as service mobility, data offloading and disruption tolerance so it can be estimated if the current load in the system can fit the new slice.

Sciancalepore~et~al.~\cite{sciancaleporeMobile2017} developed an admission control module for slice admission into a mobile network.
Their model assumes that the bottleneck of the network is the physical resource (spectrum) which is to be shared among the network tenants.
The information provided by the tenant to InP at the slice request include maximum resource utilization, duration of the slice and traffic class.
In this context, traffic class specifies some behavior of the traffic, i.e., delay tolerance and if the bit rate should be guaranteed or not.
The work considers a total of 6 traffic classes.
Once deployed into the system, tenants request resources according to a Poisson process and the InP must provide them, otherwise a Service Level Agreement (SLA) violation penalty is incurred.
The solution proposed to solve this problem applies a prediction of the traffic load of the requested slice.
Based on this and the predictions of the previously admitted slices, the admission module can evaluate if the new slice can be placed into the system.
The combination of all the possible slices in the system is modeled as a geometric knapsack problem.
When the slice leaves the system, the prediction module (as part of the admission) is informed of the actual behavior of the slice so it can evaluate how accurate its prediction was and update its knowledge with new experience.

Another system for slice admission is studied by Bega~et~al.~\cite{begaOptimising5GInfrastructure2017}.
Their model of mobile network has physical resource (spectrum) as bottleneck and they have two types of traffic classes: elastic and inelastic.
Inelastic users are characterized by having an SLA which specifies that all the requested resources must be provided when needed.
Elastic users, on the other hand, do not require a specific number of resources and can cope with a variation of the number of allocated resources and their SLA is specified by an average resource availability.
A slice request is composed of the slice duration, the traffic type and the slice size (in number of clients).
The admission problem is modeled as a Markov Decision Process (MDP).
The states are the number of elastic and inelastic users.
The actions are: accept or reject the slices, while the objective is to admit as many slices as possible while guaranteeing the tenants requested SLA.
Bega~et~al.~\cite{begaOptimising5GInfrastructure2017} proposes the use of Q-Learning to solve the problem.
They compare their solution with two heuristics and an analytical algorithm.
They show that their proposed solution can adapt if the system does not behave as modeled and provide better decisions than the other proposals.

Apart from the research works targeting new approaches to network slicing and resource management, there is significant work being done on developing platforms which can be used to integrate such solutions in real networks.
The seminal work on the platform side was started with the European Telecommunications Standards Institute (ETSI) NFV group that released a whitepaper outlining how network infrastructure made of physical nodes would be transformed to a software system running on general purpose servers~\cite{attNetworkFunctionsVirtualisation2012}.
Subsequently, the group presented the NFV Management and Orchestration Framework (MANO) that has been the reference architecture for many platforms currently being developed for virtualized network management~\cite{dahmen-lhuissierOpenSourceMANO2018}.
The reference implementation of the ETSI MANO architecture is called Open Source MANO (OSM) and is actively maintained by the open source community.
Similar initiatives were started by vendors and commercial entities to produce carrier grade options resulting in platforms like Open Orchestrator Project (OPEN-O); Enhanced Control, Orchestration, Management \& Policy (ECOMP), and more recently their converged realization Open Network Automation Platform (ONAP)~\cite{onapONAPArchitectureOverview2018}.

Although work on the development of MANO platforms is ongoing together with standardization on the architecture, interfaces and functionality, the need for developing intelligent solutions for core features like network slicing remains with the platforms providing an easier path towards integrating them in a realistic environment.
In this work we address this issue and present a policy agent for slice admission control in virtualized 5G networks.

\section{Network Slice Admission Control}\label{sec:slice-admission}
This section provides an overview of a high-level network management loop that can be applied at multiple levels of the network, specially to control the admission of new slices.
We present the system model considered by this paper, as well as our proposal to solve the admission problem.

\subsection{The Control Loop}\label{subsec:control-loop}
\begin{figure}
    \centering
    \includegraphics[scale=0.3]{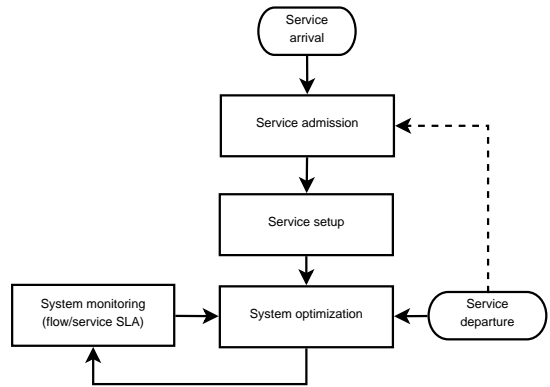}
    \caption{Flowchart representing an overview of the general network management loop.
    The dashed line represents exchange of information, for example,
    reporting the overall satisfaction experienced by the service during its lifetime.}\label{fig:control-loop}
\end{figure}

Network management operations in many levels of the network such as service admission or orchestration can be modeled as a closed loop operation, as depicted in Fig.~\ref{fig:control-loop}.
In general, a service arrives at some part in the network which then goes through an admission policy that decides whether it is in the interest of the network operator or infrastructure provider to accept the service or reject it.
Services that are admitted into the network require some setup. For example, the service might consume some infrastructure resources, and the network may have multiple resource pools from which the resources could be provided, so the decision of which pool to use is taken in this step.
Once the new service is deployed, the system enters in a general control loop where not only the deployed service instances, but also the network, is constantly monitored and optimized.
The closed loop monitoring and control mechanism ensures that if a deployed service instance needs more or less resources than it did at the deployment time, then those additional resource requirements can be accommodated or unused resource be allocated to another slice in the network.
Additionally, in the case of service completion or departure from the network, the closed loop operation ensures that not only the resources are taken away from the departing service but also the state of remaining services is optimized.
For an intelligent admission policy, the SLA parameters observed by the departing service during its lifetime are extremely important,  since a positive or negative feedback can be used by the admission policy to optimize its performance for future decisions.
The management and control loop depicted in Fig.~\ref{fig:control-loop} can be adapted to operate in other context and various parts of the network.
To demonstrate this general management concept, we will apply it to our work proposal where we use the closed loop approach to a high-level network operation in which network slices are deployed onto a shared network infrastructure.

In this network sharing context, there are three distinct roles comprising the InP: the entity that owns the infrastructure on which the slices will be executed; the tenant, entity that requests resources from the InP to run its services;
and a user, which will consume the services from the tenant.
A slice deployment request made from the tenant contains a SLA requirement and the InP must provide enough resources to fulfill the SLA.
Examples of SLA include network coverage over a certain area and minimum network bandwidth.
The InP has limited resources and, therefore, in some situations cannot fulfill the SLA for all the tenants requesting slices.
That motivates the existence of the slice admission module.
Its objective is to admit as many slices as possible into the system, with the objective of maximizing resource utilization, consequently increasing the revenue for the infrastructure provider.
The constraint is that it should not allow slices that would have their SLA violated, or cause SLA violation for the other deployed services.
A network slice may require resources in multiple parts of the network.
For example, processing power in base stations or connectivity in the backhaul.
To setup those resources, certain decisions must be made.
For example, there are usually multiple paths connecting the base station to the core network, and the decision of which of those paths will provide the required connectivity for a particular slice is done at the service setup module.

Once the slice is admitted into the network, it becomes operational and the system enters in its main control loop with respect to that service.
Resources in the system are constantly optimized so that their allocation to each slice matches the real-time service needs.
The system is also constantly monitored, so that observed SLA by all the deployed slices is recorded and evaluated for compliance.
The last significant change in the system happens when the lifetime of a slice ends, and it must leave the network.
This triggers an optimization of the system, so it can optimize its resources to the slices currently deployed.
The departure of a slice is also reported to the admission control, so that it can evaluate the consequences of the admission of other slices as well as the slice behavior. The latter enables the admission control to learn how to make better admission decisions in future.

\subsection{System Model}\label{subsec:system-model}
In this section we present the details of the considered system model for the proposed neural network-based policy agent for slice admission control.
Our system model considers not only physical resources, but also the high-level components of the network such as Virtual Network Functions (VNF) and computational and connectivity resources that can all be a bottleneck for the different slice classes in the network.
As the main objective is to evaluate the policy agent for slice admission against other approaches, our system model is based on the concepts developed in Raza~et~al.~\cite{razaSliceAdmissionPolicy2018a}.

\begin{figure}
    \centering
    \includegraphics[width=\textwidth]{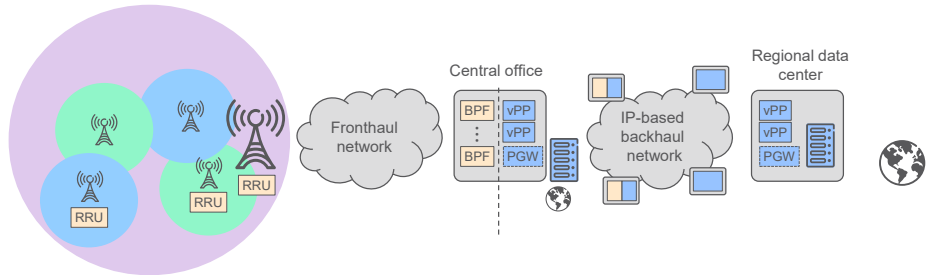}
    \caption{Overall architecture in a flexible mobile network.
    Slice require functions that can be placed and
    consume resources in different parts of the network.}\label{fig:overall-arch}
\end{figure}

The overall network architecture is presented in Fig.~\ref{fig:overall-arch}.
Its main components, modeling a metropolitan area, are: a couple of Regional Data Centers (RDC), a few dozes of Central Offices (CO), and hundreds of Remote Radio Units (RRU).
The RDC provides connectivity to external networks and has General Purpose Processors (GPP).
The CO has both the Special (radio) Purpose Processor (SPP) and GPP.
Those resources at the CO are more expensive than in the cloud but can deliver a lower latency.
The RRU provides the radio access to the end users.

At the assumed abstraction level, there are some components that can be deployed by slices.
The edge (CO) and core (RDC) both provide GPP, which can be used by slices to execute general processing functions such as a virtual Package Processor (vPP); mobile network functions, such as Package Gateway (PGW); and slice specific applications.
The other resource is connectivity, which enables communication between edge and core.
Each CO $c \in \mathcal{C}$ has a capacity of $g_c$~GPP, each RDC $r \in \mathcal{R}$ has a capacity of $g_r$~GPP and each link $l \in \mathcal{L}$ has a capacity of $d_l$~link capacity units.
In the assumed model, all the resources have integer units.

The described resources are consumed by slices that deploy the presented components in the network.
The maximum number of GPP that a slice can request at each CO is $k_c$, and at any RDC is $k_s$;
and the maximum number of connectivity resource between them is $k_m$.
Deployed slices have dynamic resource requirements over time and if the requested resources cannot be provided by InP, an SLA violation occurs.

Tenant $t$ requesting a network slice must inform InP of their immediately requested GPP at each CO $j_{c}$ (which indicates the region it wants to have coverage in); the number of GPP at a RDC $j_{s}$; the connectivity between them $j_{m}$; the duration $j_e$; and the priority $j_p$.
If an InP reserves enough resources for the admitted slice, it is guaranteed that no SLA penalty will be imposed and the tenant will remain satisfied.
However, the tenant does not always use the maximum number of allocated resources and reserving them leads to resource underutilization.
Therefore, the InP can try to understand the behavior of its tenants and sometimes oversubscribe the system by deploying additional slices so that, with managed risk of causing SLA violation, a higher revenue is achieved.

When the slice ends its service life cycle, the InP receives its revenue for hosting the service which is a fixed amount agreed at the time of admission based on slice parameters.
Any SLA violation causes a decrease on this value which is proportional to the magnitude of the violation.

The described environment model allows for the study of slice admission and its consequences in the system.
In the next section,
 it will be used to study new techniques for training Reinforcement Learning (RL) slice admission agents.

\subsection{Slice Admission by Reinforcement Learning}\label{subsec:rl-admission}

In the network slicing scenario, the objective of the InP is to increase the revenue.
In the general management loop presented in Section~\ref{subsec:control-loop}, one of the decisions that can be optimized is the slice admission.
Assuming that all the other procedures (setup, scaling, etc.) are established, we study a RL agent that will learn when to accept or reject a slice into the system to maximize profit.
Such an agent would have to learn how the system behaves so that it can consider the current load in the system, understand the risk of causing an SLA violation and decide on the slice admission.

\begin{figure}
    \centering
    \includegraphics[scale=0.3]{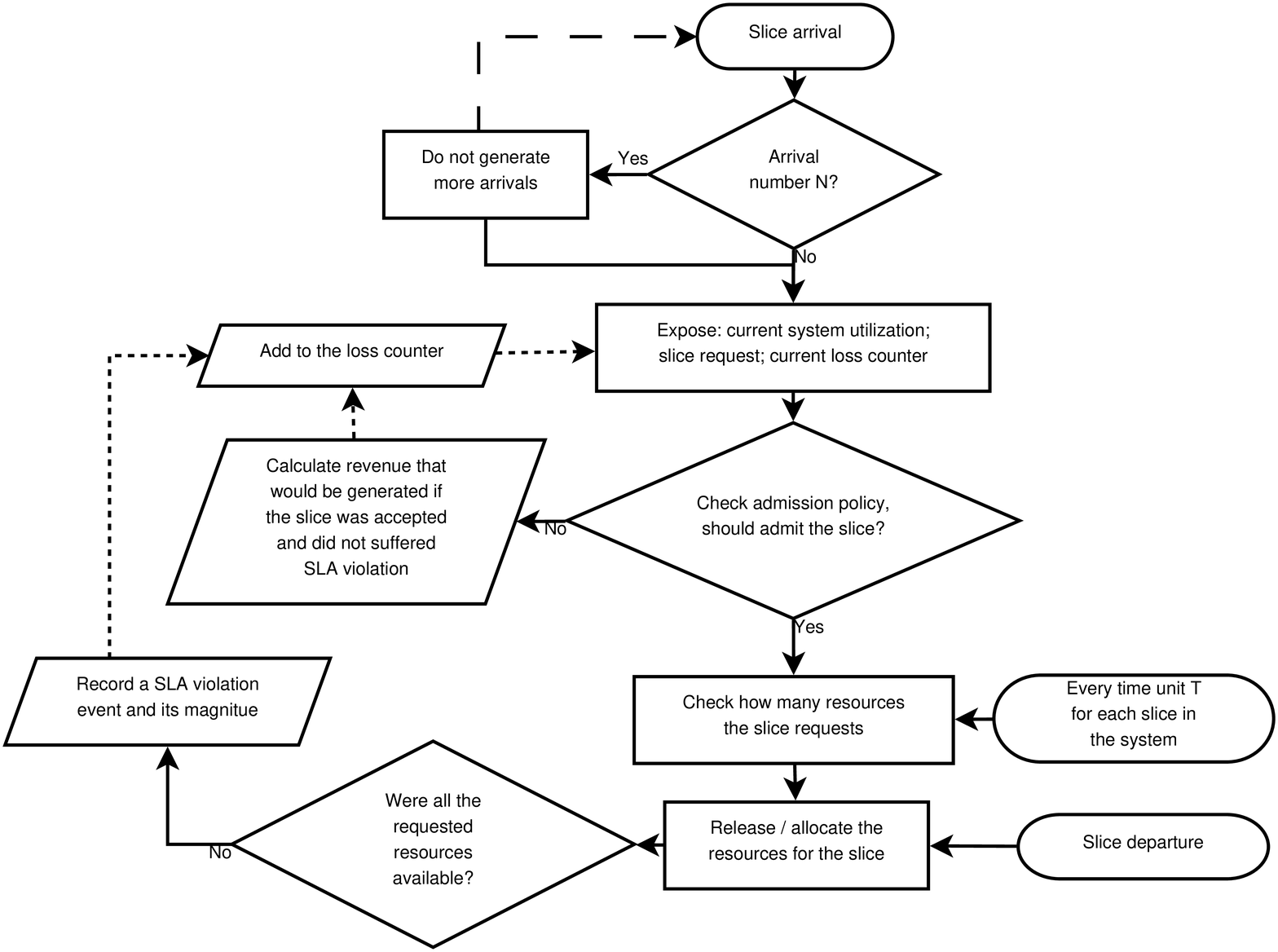}
    \caption{Overview of the experience gathering in the network slicing system.}\label{fig:experience-collection}
\end{figure}

The overall modeling of the experience acquisition in the network slicing system is presented in Fig.~\ref{fig:experience-collection}.
The diagram emphasizes the data that will be generated and on which data the agent will learn from.
In the system, there are three events: slice arrival, slice departure and a periodic check of slice health (requested resources).

To make the learning suitable for RL, the system is modeled in an episodic fashion.
Each episode is determined by the arrival of $N$ slices.
The objective is to accept as many of those as possible, while avoiding SLA violation.

Upon a slice arrival, the agent is consulted to make a decision.
It is made aware of the current system utilization, the slice request parameters and the accumulated loss suffered by the system until the time of arrival.
Based on that, it either accepts or rejects the slice.
If rejected, the system interprets that some revenue could be obtained if the system had resources to admit the slice thus the rejection is interpreted as a loss.
Accepted slices are allocated the number of resources requested, if those are available, if only a fraction of the resources is available, those are allocated, and an SLA violation is recorded.

The slice scaling event revisits all deployed slices in the system and checks how many resources the slices are requesting and adjusts the resource allocation.
As before, if only a fraction of the requested resources is available, the fraction is allocated and an SLA violation is recorded.
The scaling event is fired periodically with a period $T$.

When the slice finishes its execution, a slice departure event is generated, which releases resources allocated to the slice immediately.
When all the $N$ slices arrive and the accepted ones finish their execution, an episode is finished.

Aligned with previous work~\cite{razaSliceAdmissionPolicy2018a,maoResource2016}, our RL agent is using a policy network, which has a configurable number of inputs and hidden layers and two outputs, representing the probability of accepting and the probability of rejecting the slice.
The input to the neural network is: the system utilization, the amount of each resource requested by the slice, its duration, priority and tenant.
This information is encoded in a binary representation according to $\mathcal{A}_b(x)$, which creates an bit field of $b$ bits with the first $x$ bits equal to one and the others $b-x$ bits equal to zero.
The state vector $s$, when slice $j$ is requesting to enter the system, is then created by concatenating the bit fields: $\mathcal{A}_{g_c}(n_c)$, $\mathcal{A}_{k_c}(j_c)$, $\mathcal{A}_{g_r}(n_r)$, $\mathcal{A}_{k_s}(j_s)$, $\mathcal{A}_{d_l}(n_l)$, $\mathcal{A}_{k_m}(j_m)$, $\mathcal{A}_{k_e}(\min(j_e, k_e))$, $\mathcal{A}_1(j_p)$ and $\mathcal{A}_{n_t}(t)$, where $n_c$, $n_r$ and $n_l$ are the number of busy resources at $c$, $r$ and $l$, respectively, $c \in \mathcal{C}$, $r \in \mathcal{R}$, $l \in \mathcal{L}$, and $k_e$ represents the maximum requested duration observable by the agent.

\section{Evaluation}\label{sec:evaluation}

\begin{figure}
    \centering
    \includegraphics[scale=1]{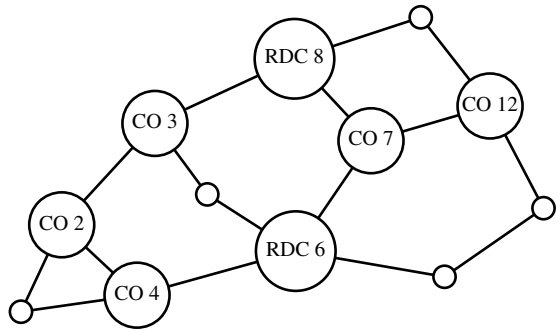}
    \caption{Network topology used to evaluate the system.}\label{fig:topology}
\end{figure}

We evaluate the proposed method using a network topology~\cite{razaResourceOrchestrationMeets2018} shown in Fig.~\ref{fig:topology}.
The nodes with a high degree of connectivity were selected as RDC, some with medium degree as CO, and some nodes are connection points that provide connectivity routes.

Each RDC has a capacity $g_r=80$, each CO $g_c=50$ and each link $d_l=50$.
The reference behavior of slices (profiles) are the ones reported by Raza~et~al.~\cite{razaPriorityAwareServiceOrchestration2017}.
The resource requirements of each slice are given by the time of the day and the type of the slice.
Between 9:00 and 19:00 the high-priority slice requests 20 GPP resources at the CO, 5 connectivity resources and 5 GPP at the RDC, this is usually the busy hours at business districts where high-priority traffic is likely to occur.
Other than at those hours, the high-priority slice requires only 15 GPP at the CO.
The low-priority slice between 16:00 to 22:00 requires 10 GPP at the CO, 10 connectivity resources and 10 GPP at the RDC.
This is likely to be the busy hour in residential districts, where low-priority slices are likely to exist. During the other hours,
low-priority slices require 5 GPP at the CO, 5 connectivity resources and 5 GPP at the RDC.
This configuration reproduces a network that was projected to have as main bottleneck the resources at the CO, which are the most expensive.

In contrast with previous work, which considered fixed slice profiles, we assume that the resource requests of a slice is non-deterministic.
The current simulator accepts integer resource requests.
Consequently, we chose to use a Binomial distribution with 5 trials and a tenant-dependent success probability ($u_t$) to generate a \emph{noise}.
This noise is sampled every time unit and subtracted from the resource request of each resource type of the reference profile, if this subtraction leads to a negative resource request, it is understood as no resource needed (0).
The slices impose penalties as discussed in Section~\ref{sec:slice-admission} and the magnitude of those penalties is proportional to the tenant penalty weight ($v_t$).

We configured the system to run a simulation with two tenants, $t=0$ and $t=1$, and used $u_0=0.1$, $u_1=0.9$, $v_0=1$ and $v_1=0.1$.
This setup makes tenant 0 have a higher network usage, and his penalty is higher than for tenant 1, which imposes a lower usage.

We consider as baseline an admission agent that is not aware of the tenant who is making the request~\cite{razaSliceAdmissionPolicy2018a}.
The proposed strategy adds as input to the admission agent the ID of the tenant who is requesting the slice, as described in Section~\ref{subsec:rl-admission}.
Both agents are using a policy network with 4 hidden layers with 40 neurons in each and a ReLu activation function, following the baseline.

We trained the system for 10,000 iterations, with 25 episodes per iteration, each episode with a load of 80 Erlangs and 600 arrivals.
With the trained model we ran the test for a new set of 25 episodes with the same configuration.

Alongside our proposal (Prop), we show results for four other policies.
BL is the baseline, which was adapted from Raza~et~al.~\cite{razaSliceAdmissionPolicy2018a}. With the exception of the tenant ID, BL has the same input as Prop.
RND is the random police, i.e., it chooses to accept or reject with equal probability.
Fit accepts the slice if the InP has enough resources to fulfill its request at admission time.
Those heuristics were defined and also used by Raza~et~al.~\cite{razaSliceAdmissionPolicy2018a}.

\begin{figure}
    \centering
    \includegraphics[scale=1]{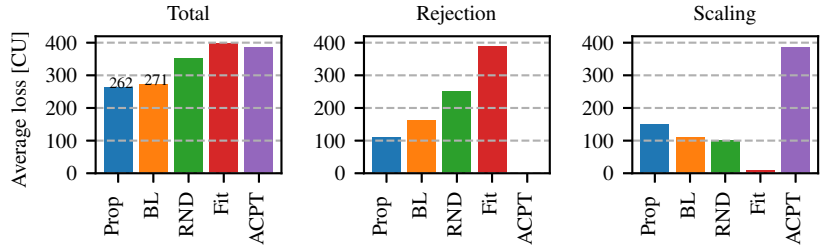}
    \caption{Overall results. Prop is the proposed policy, BL is the baseline, RND is random,
    Fit accepts a slice if there are enough resources at admission time, and ACPT is accept all the slices.
    The left graph shows the overall loss achieved by each policy, BL has a loss of 271, while Prop has 262.
    In the middle, only the loss incurred by rejecting the slice is shown and, on the right,
    only the loss incurred by scaling.}\label{fig:overall-results}
\end{figure}

The results are summarized in Fig.~\ref{fig:overall-results}.
We can observe that accepting all the slices incurs a high scaling loss which signals that some of the slices could have been rejected.
Fit is too conservative and rejects too many slices causing a resource underutilization.
Randomly accepting the slices essentially accepts half of them.
The baseline learns a better policy and achieves a balance between rejecting some slices and handling some scaling loss.
However, it does not have information on which tenant is requesting the slice.
Thus, it cannot achieve the performance of the proposed solution which can fine-tune the decision to the specific tenant, and consequently find a better balance.

We suppose that the baseline can still infer the tenant by the level of the usage which is indeed a function of the tenant (given $u_t$) and is present at its input.
However, because the resource usage is noisy, it probably cannot achieve the best possible information about the tenant which is available for the proposed policy.
To better understand which services the policies are choosing, we analyze the rejection probability for each class of slices.

\begin{figure}
    \centering
    \includegraphics[scale=1]{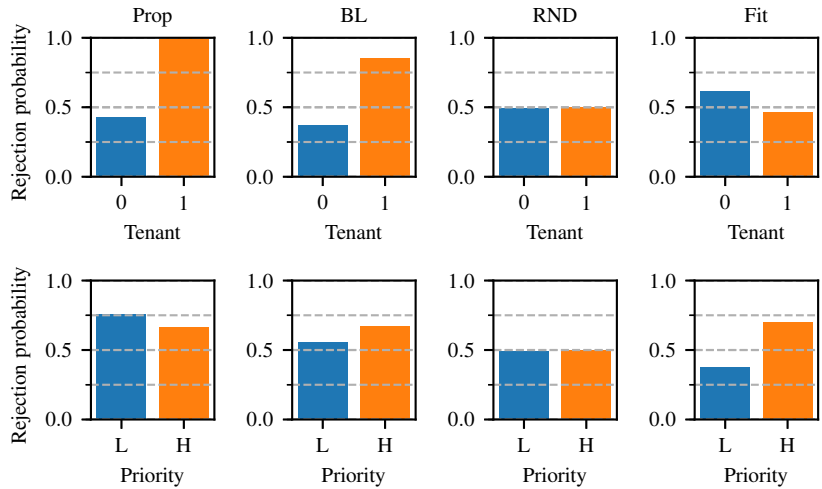}
    \caption{Rejection probability for each policy (columns), marginalized by tenant (top line), or by priority (bottom line). ACPT policy is not shown, because it is zero for all cases.}\label{fig:rejection-tenant-or-priority}
\end{figure}

We analyze which slices are being rejected in Fig.~\ref{fig:rejection-tenant-or-priority}.
We can see that Fit rejects more slices of tenant 0.
That happens because tenant 0 usually requests more resources compared to tenant 1 (given that $u_0<u_1$) and so it is more probable that there are enough resources for its slices.
Random accepts half in any marginalization by nature.
The baseline accepts more slices from tenant 0.
It should have learned that rejecting tenant 0 slices incurs a higher penalty ($v_0>v_1$) but instead it was probably inferred from the resource.
However, the balance found by the baseline rejects more high-priority slices than low-priority ones, which is counter intuitive, yet this can be due to no tenant awareness, for example, the configuration of the system makes off peak hours slices from tenant 0 and peak hours slices from tenant 1 have similar resource usage and priority, but different penalties.
Finally, the proposed strategy seems to reject almost all the tenant 1 slices in exchange for a higher acceptance of tenant 0.
It also manages to find a point where higher priority slices are more often accepted compared to lower priority ones.

\begin{figure}
    \centering
    \includegraphics[scale=1]{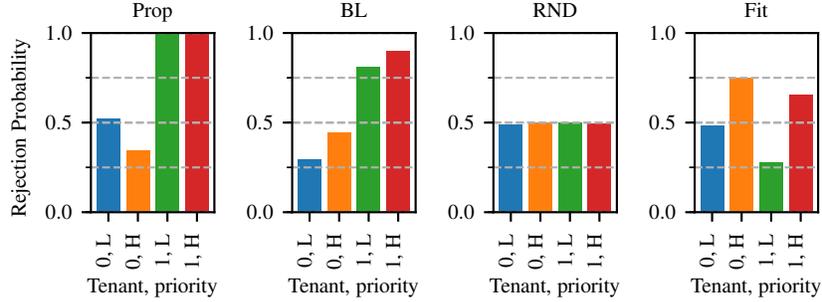}
    \caption{Rejection probability for each policy (columns) marginalized by tenant and priority.
    Accept all is not shown, since it is zero for all the cases.}\label{fig:rejection-tenant-and-priority}
\end{figure}

Another way to investigate how slices are being classified is to examine the rejection probability marginalized by tenant and priority, shown in Fig.~\ref{fig:rejection-tenant-and-priority}.
We can use the same rationale for Fit: it accepts more of the low priority from tenant 1, then low priority from tenant 0, then high priority from tenant 1 and high priority from tenant 0.
That is exactly the expected sequence of slices if they are ordered by the expected magnitude of its resource request when compared to each other.
Random once again accepted half of each class.
For the baseline, we can see that it understands that some classes are more important than others, but it seems, for example, that it cannot identify that high-priority slices from tenant 0 are more valuable than low priority ones from the same tenant.
That is one of the behaviors that the proposed method could learn: it also rejects most of the slices from tenant 1, that can happen because the load is high, and there are enough slices from tenant 0, so it can be lucrative enough, at lower loads the behavior can be different.

\section{Conclusion}\label{sec:conclusion}

This work presented a reinforcement learning-based policy agent for slice admission control in virtualized 5G networks.
We evaluated our proposal in a scenario where slice requests have a stochastic resource requirements footprint and admission control is non-trivial.
Our network model included the concept of different tenants that give different revenues, consequently increasing the possible combinations of admission decisions.
In this context, we showed that the tenant-awareness contributes to a better admission policy and brings more revenue to the InP.
We evaluated our proposed admission policy in a simulated environment and compared the performance to other related strategies.
As future work, we intend to evaluate the policy agent in more challenging system context where network and slices dynamics are increased.
Another topic to be investigated is the proportion of tenants.
The presented results only considered a uniform number of requests for each tenant but in real networks, the number of slices deployed or requested for each tenant will change over time.
Finally, the performance of other machine learning algorithms in a similar dynamic system will be evaluated.
%
%
\bibliographystyle{splncs04}
\bibliography{zotero,additional}

\end{document}